\documentclass{andromedaone}

\received{xx January 2018}
\published{xx March 2018}

\def\be{\begin{equation}}
\def\ee{\end{equation}}
\def\bea{\begin{eqnarray}}
\def\eea{\end{eqnarray}}

\usepackage{lineno}
\usepackage{graphicx}%

\begin{document}

\title{Summary of CMS Higgs Physics}

\author{Walaa Elmetenawee\auno{1,2} on behalf of the CMS collaboration}
\address{$^1$INFN - Sezione di Bari, Via Giovanni Amendola, 173, 70126 Bari, Italy}
\address{$^2$Physics Department Faculty of Science, Helwan University, Ain Helwan 11795 Cairo, Egypt} 
\begin{abstract}

Since the discovery of the Higgs boson in 2012, substantial advancements have been achieved in exploring its characteristics. The utilization of extensive data sets has facilitated recent results, enabling not only the determination of the Higgs boson mass and total production cross section in the most sensitive decay channels, but also measurements of fiducial and differential cross sections, as well as searches for rare or exotic processes. The study of the Higgs boson pair production which is fundamental to the study of the Higgs boson self-coupling, received a significant boost, too. These proceedings focus on the latest Higgs physics results achieved by the CMS Collaboration using the entire dataset collected during Run-2 of the LHC, corresponding to an integrated luminosity of approximately 140 fb$^{-1}$.
\end{abstract}

\maketitle


\section{Introduction}
In July 2012, the ATLAS and CMS Collaborations~\cite{Atlas, CMS} at the CERN LHC jointly announced the discovery of a scalar particle consistent with the predicted standard model (SM) Higgs boson~\cite{HiggsAtlasDis,HiggsCMSDis,HiggsCMSDis_2}. Following this landmark discovery, both collaborations began a detailed exploration of the particle's properties, clearly demonstrating its CP-even structure and measuring its mass to be compatible with 125 GeV. Since the SM does not predict the precise value of m$_{H}$, its experimental determination becomes crucial to understand the accessible production modes and decay channels of the Higgs boson at the LHC. Leveraging the datasets from Run 2 of the LHC, precise measurements of the Higgs boson properties are within reach. The mass and width are known with uncertainties on the order of MeV, and most of the major production mechanisms and decay channels have been experimentally observed. Furthermore, the available dataset is large enough to facilitate fiducial and differential measurements, crucial for a deeper understanding of the SM and the exploration of potential deviations that could hint at physics beyond the SM (BSM). Employing dedicated analysis techniques, sensitivity to the Higgs boson coupling to second-generation fermions was obtained, CP properties were probed, and substantial advances were made in understanding di-Higgs boson production.

\section{Higgs boson mass and width }

\begin{figure}[ht]
\centering
\includegraphics[width=15cm] {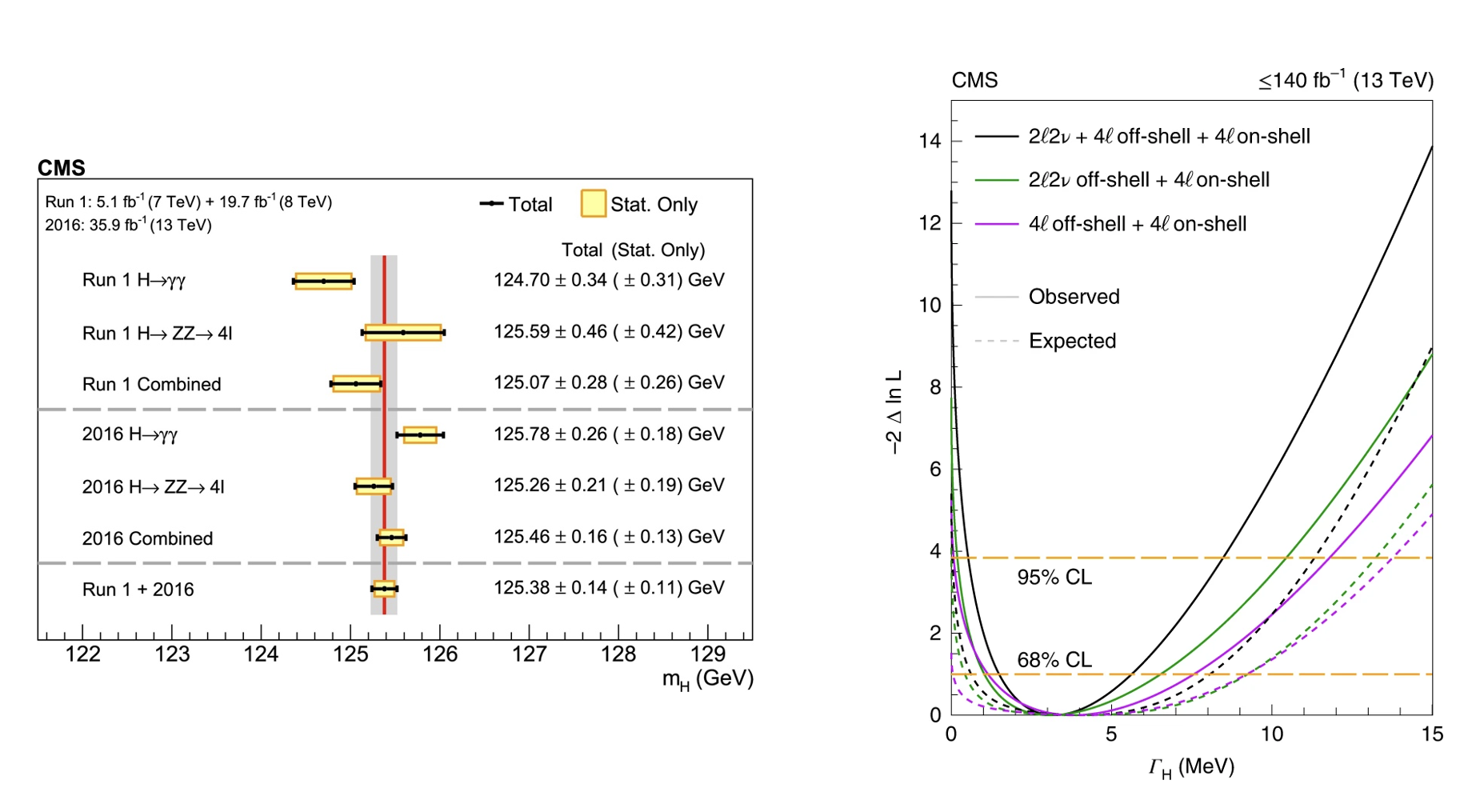}
\caption{Left: A summary of the measured Higgs boson mass in the $\textnormal{H}\rightarrow\gamma\gamma$ and $\textnormal{H}\rightarrow ZZ\rightarrow4\ell$ decay channels, and for the combination of the two is presented~\cite{Mass}. Right: observed (solid) and expected (dashed) one-parameter likelihood scans over the Higgs width $\Gamma_{H}$. Scans are shown for the combination of $4\ell$ on-shell data with $4\ell$ off-shell (magenta) or $2\ell2\nu$ off-shell data (green) alone, or with both datasets (black)~\cite{Width}.}
\label{Mass_Width}
\end{figure}

The Higgs boson mass is measured through the two discovery channels ($\textnormal{H}\rightarrow\gamma\gamma$ and $\textnormal{H}\rightarrow ZZ$), chosen for their optimal invariant mass resolution and a good signal-to-background ratio around the Higgs boson mass peak. Accurate energy and momentum calibrations, relying on meticulous detector calibration and alignment, are crucial for these analyses. The most precise result from CMS at the time of the conference~\cite{Mass}, which combines data from 2016 with those from Run 1, yielding a Higgs boson mass of m$_{H}$ = 125.38 $\pm$ 0.14 GeV, with an uncertainty of 0.11\%. Fig.~\ref{Mass_Width} (left) illustrates a summary of the individual and combined measurements using the Run 1 and 2016 datasets. Since these results were presented, a more precise measurement in the $\textnormal{H}\rightarrow ZZ\rightarrow4\ell$ channel was made public by the CMS Collaboration (see this Ref.~\cite{Mass_new}).

Direct measurements of the Higgs boson mass lack sensitivity to its intrinsic width, given its significantly smaller scale compared to the typical invariant mass resolution of the CMS experiment. Indirect measurements exploit the proportional relationship between off-shell and on-shell productions of a resonance and its width:
\begin{equation}
\label{eq:1}
\sigma^{\textnormal{on-shell}} \propto \frac{g_{p}^{2}g_{d}^{2}}{\Gamma_{\textnormal{H}}}\propto \mu_{p}, \sigma^{\textnormal{off-shell}} \propto g_{p}^{2}g_{d}^{2} \propto \mu_{p} \Gamma_{\textnormal{H}}
\end{equation}
where $g_{p}$ and $g_{d}$ are the couplings associated with the Higgs boson production and decay modes, respectively, and $\mu_{p}$ is the on-shell Higgs boson signal strength. CMS pursued this approach using the full Run 2 dataset and the $\textnormal{H}\rightarrow ZZ$ final state~\cite{Width}, revealing evidence of off-shell Higgs boson production. Fig.~\ref{Mass_Width} (right) illustrates likelihood scans across $\Gamma_{\textnormal{H}}$. These scans consistently incorporate information from the $4\ell$ on-shell data. The three presented cases involve the inclusion of the $4\ell$ off-shell data alone, the $2\ell2\nu$ off-shell data alone, or both. The log-likelihood curves exhibit steep slopes near $\mu_{\textnormal{off-shell}}$ = 0 and $\Gamma_{\textnormal{H}}$ = 0 MeV due to interference terms between the H-boson and continuum-ZZ production amplitudes. The measured Higgs boson width, $\Gamma_{\textnormal{H}} = 3.2_{-1.7}^{+2.4}$ MeV, is compatible with the standard model's expected value of 4.1 MeV.

\section{Cross section measurements}

Inclusive and differential measurements of the Higgs boson production cross section are important tools to test the SM expectations and to probe BSM physics. Simplified template cross sections (STXS) measurements and fiducial differential cross section measurements have been performed in the $\textnormal{H}\rightarrow\gamma\gamma$~\cite{STXSHgaga, FidXSHgaga}, $\textnormal{H}\rightarrow \textnormal{ZZ}$~\cite{STXS_fidHZZ},  $\textnormal{H}\rightarrow \textnormal{WW}$~\cite{XS_HWW},  $\textnormal{H}\rightarrow\tau\tau $~\cite{XS_H2tau}, and $\textnormal{H}\rightarrow \textnormal{b} \bar{\textnormal{b}}$~\cite{XS_Hbb} decay channels by the CMS experiment. A selection of these results is presented in this section, focusing on the measurements done in the $\textnormal{H}\rightarrow\gamma\gamma$ and $\textnormal{ H}\rightarrow \textnormal{ZZ}$ channels.

\begin{figure}[ht]
\centering
\includegraphics[width=8.3cm]{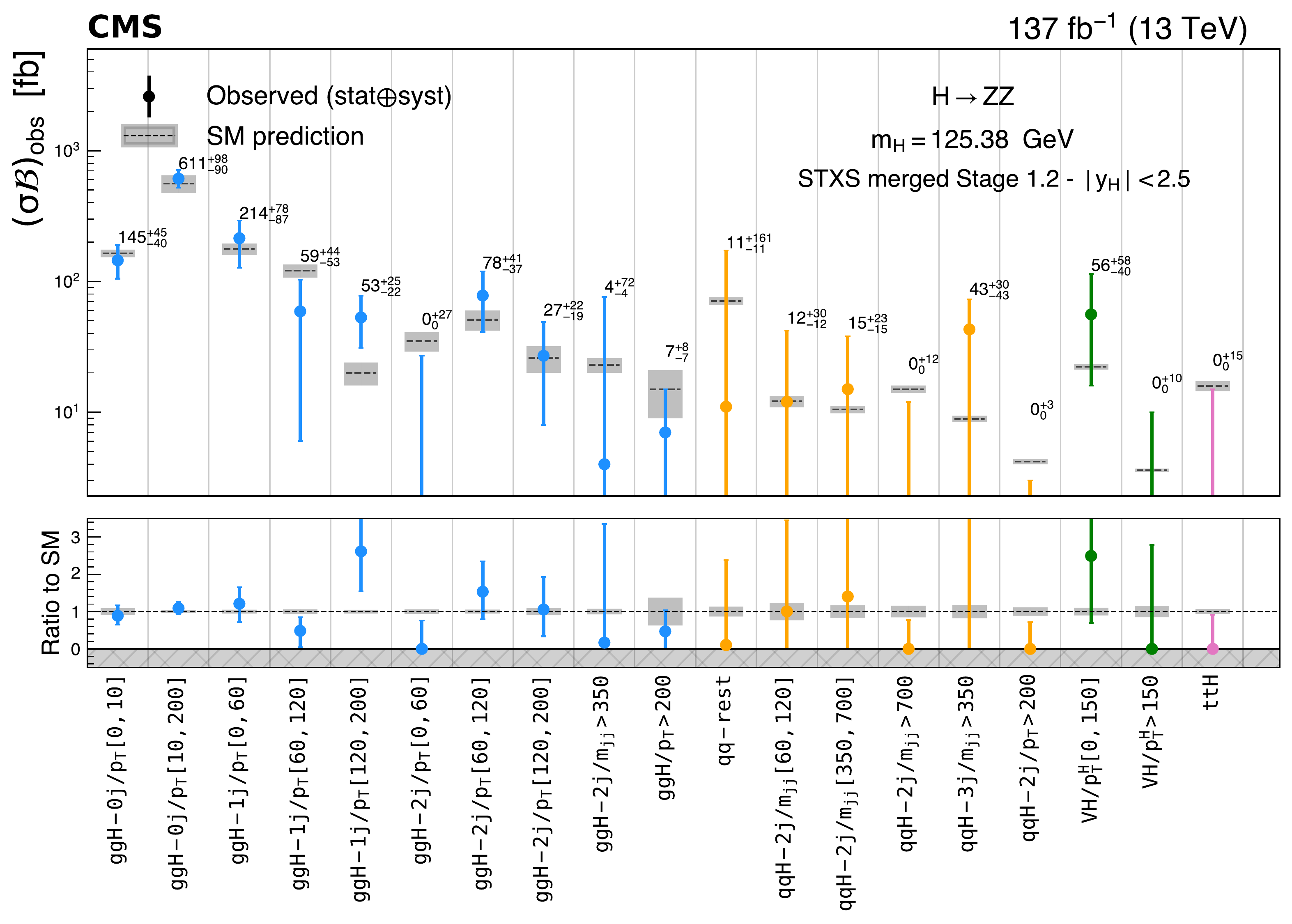}
\qquad
\includegraphics[width=8.8cm]{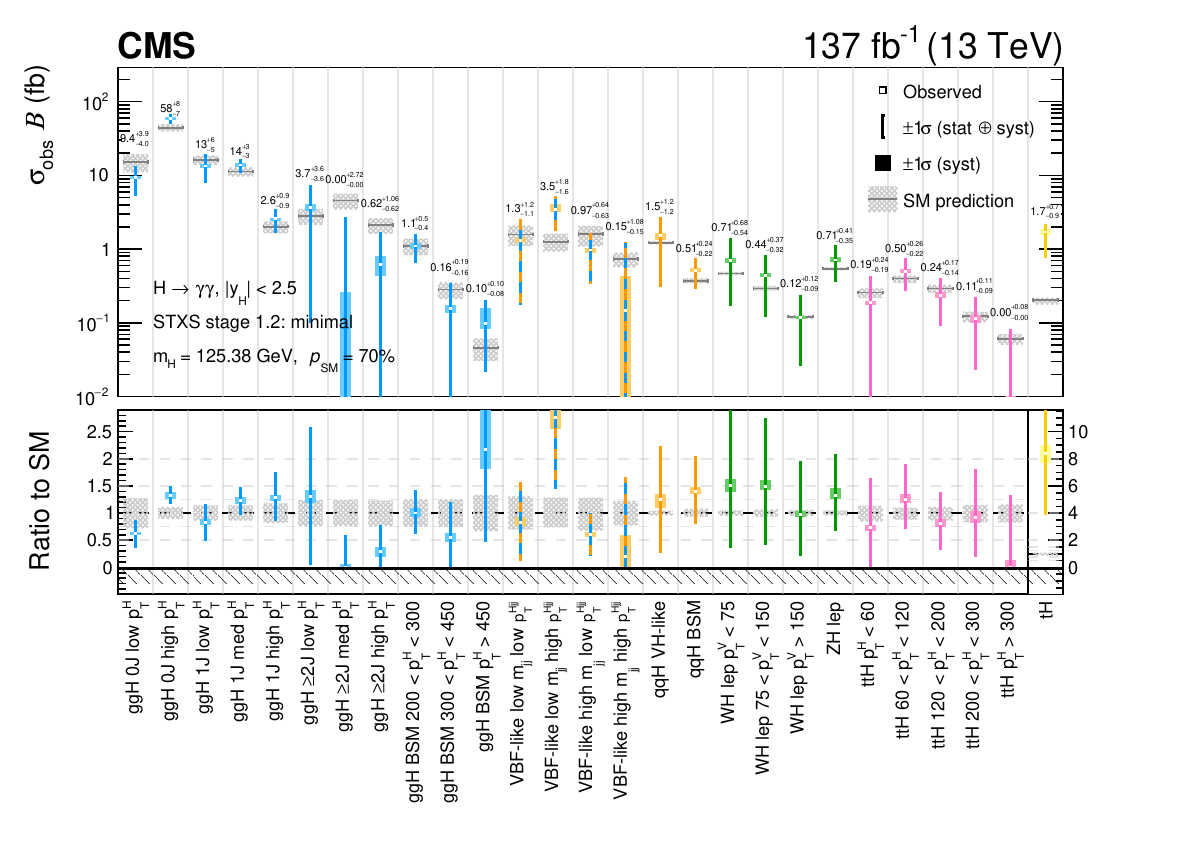}
\caption{The measured cross sections $(\sigma\mathcal{B})_{obs}$ and the SM predictions $(\sigma\mathcal{B})_{SM}$ of the merged stage 1.2 STXS production bins for $\textnormal{H}\rightarrow ZZ$ decay (left)~\cite{STXS_fidHZZ} and $\textnormal{H}\rightarrow\gamma\gamma$ (right)~\cite{STXSHgaga} at $m_{\textnormal{H}}$=125.38 GeV.}
\label{STXS}
\end{figure}

\subsection{Simplified Template Cross Sections Measurements}

STXS has been adopted as a common framework by the LHC experiments for Higgs boson measurements. It aims to enhance sensitivity to BSM physics effects while minimizing theory dependence by defining exclusive kinematic regions in the Higgs boson production phase space. In the STXS framework, there is no fiducial selection for the decay, such that measurements in different decay channels can be easily combined. The framework is organized into kinematic regions or bins, categorized in stages with increasing granularity. The most recent stage in which measurements have been performed is stage 1.2.

In the $\textnormal{H}\rightarrow \textnormal{ZZ}\rightarrow4\ell$ channel~\cite{STXS_fidHZZ}, the STXS measurements heavily utilize kinematic discriminants based on matrix-element probabilities. They are used in
the event selection, to build STXS categories, and in the 2D likelihood fit. Due to a low number of events, some bins, like ttH and the high-m$_{jj}$ VBF bin are merged. Conversely, certain ggH bins achieve a precision of up to 16\%, as shown in Fig.~\ref{STXS} (left).

The $\textnormal{H}\rightarrow\gamma\gamma$~\cite{STXSHgaga, FidXSHgaga} analysis employs Boosted Decision Trees (BDTs) and Deep Neural Networks (DNNs) to construct categories targeting STXS bins. Covering all Higgs boson production modes down to tH, the analysis leverages diphoton decay kinematics to fit the Higgs boson’s invariant mass peak over the combinatorial background. The STXS results are presented in two binning schemes: one aiming for broad STXS bin coverage (27 bins), with minimal merging to avoid excessive parameter anti-correlation, and the other adopting maximal merging until the expected uncertainty is less than 150\% of the SM prediction, as depicted in Fig.~\ref{STXS} (right). Notably, the ggH BSM bin ($p_{T}^{H} >$ 200 GeV), sensitive to BSM physics in the ggH loop, is measured with an uncertainty below 40\% and aligns with the SM prediction. Another significant result is the separate measurement of the tH production mode from ttH (usually in stage 1.2 they are measured together)and this is the best tH measurement up to date with an observed (expected) limit at 95\% confidence level (CL) of 14 (8) times the SM value.

\subsection{Fiducial differential cross section measurements}
The substantial Run 2 dataset has led to many inclusive Higgs boson cross section measurements starting to be primarily limited by systematic uncertainties, with sizable contributions from both experimental and theoretical sources. In addition to inclusive measurements, the available integrated luminosity allows for fiducial differential cross section measurements, probing interesting and previously unreachable regions of the phase space. Fiducial measurements report the cross section in bins of the observable of interest (differential) and are extrapolated to a restricted particle-level phase space (fiducial). The choice of bin boundaries is crucial and follows criteria such as alignment for combination ease, ensuring an adequate number of events for low expected uncertainty, and maintaining a good level of signal-to-background ratio. An illustrative observable is the Higgs boson transverse momentum ($p_{\textnormal{T}}^{\textnormal{H}}$), known for sensitivity to deviations in Yukawa couplings and potential BSM effects.

\begin{figure}[ht]
\centering
\includegraphics[width=7.6cm]{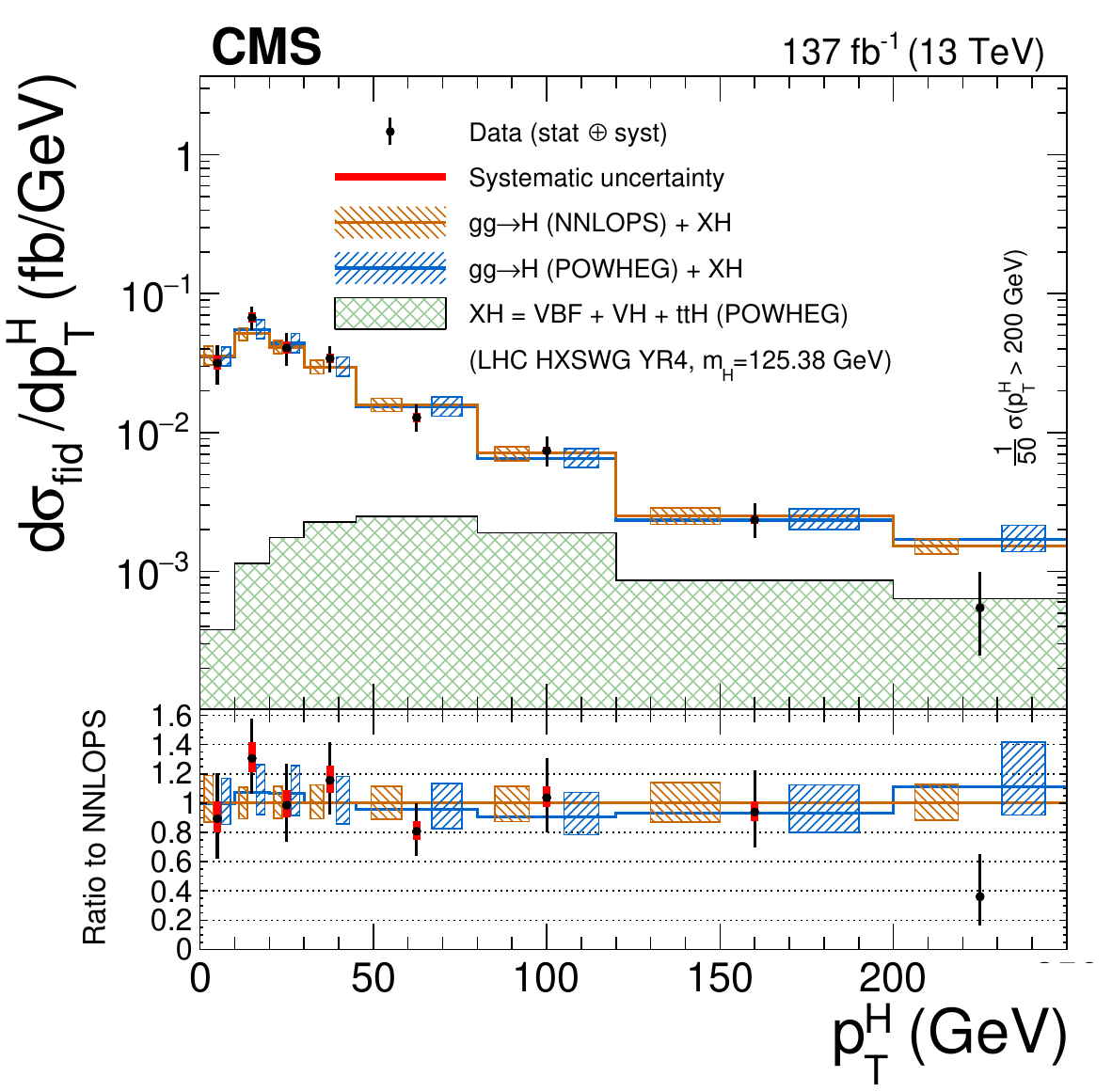}
\qquad
\includegraphics[width=7.5cm]{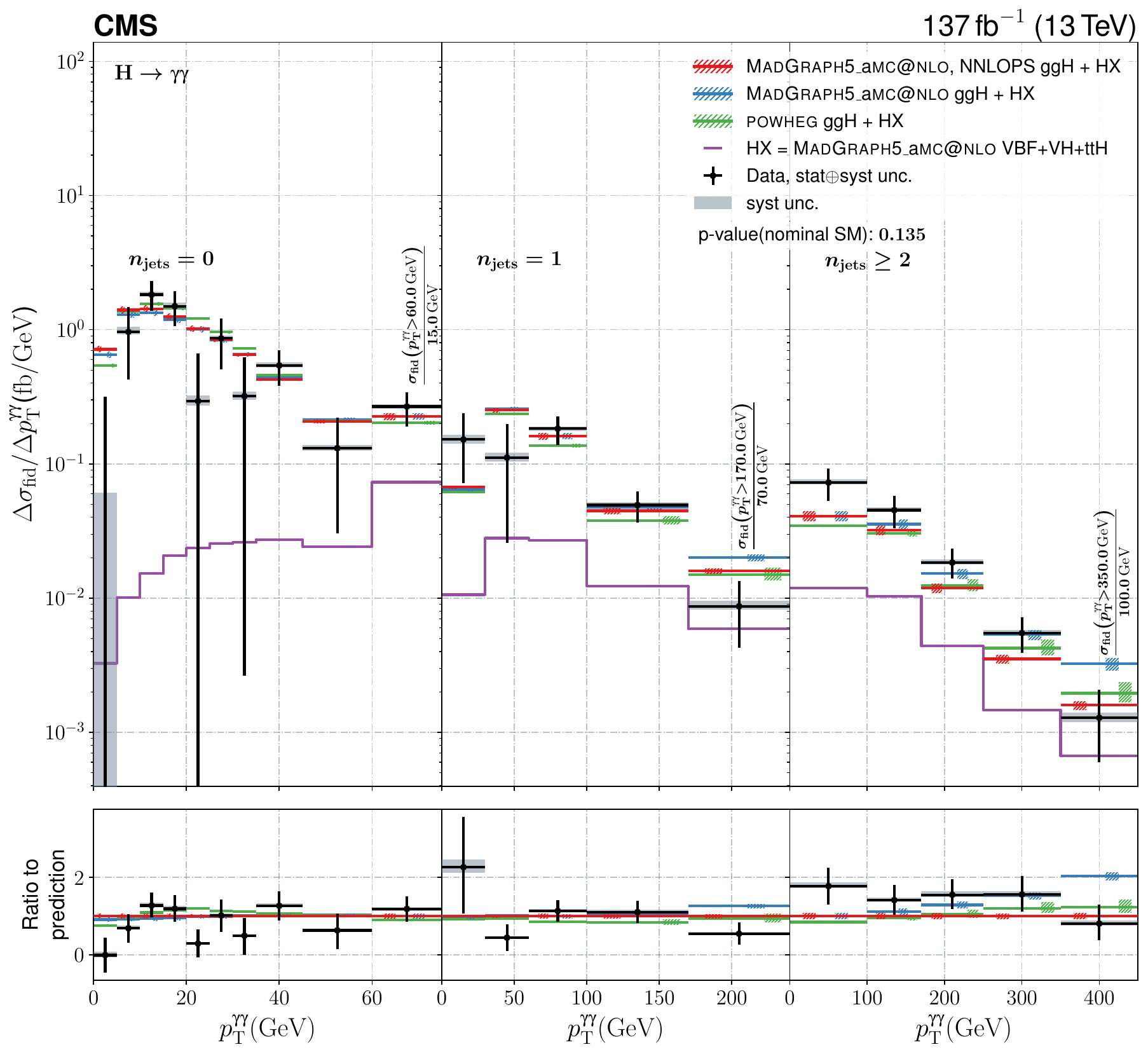}
\caption{Left: Differential fiducial cross sections measurement as a function of $p_{T}^{H}$ in $\textnormal{H}\rightarrow ZZ\rightarrow 4\ell$~\cite{STXS_fidHZZ}. Right: Double-differential fiducial cross section measured in bins of $p_{T}^{H}$ and N$_{jets}$ in $\textnormal{H}\rightarrow\gamma\gamma$~\cite{FidXSHgaga}}
\label{FidXS}
\end{figure}

In the four-lepton fiducial analysis~\cite{STXS_fidHZZ}, the integrated cross section in the three final states and the inclusive one:  

\begin{center} 
$\sigma_{\textnormal{fid}}^{\textnormal{HZZ}} = 2.84_{-0.31 (10.9\%)}^{+0.34(12.0\%)}$ fb = $2.84_{-0.22 (7.7\%)}^{+0.23(8.1\%)}$ (stat.) $_{-0.21 (7.4\%)}^{+0.26(9.2\%)}$ (sys.) fb
\end{center} 
aligns with the SM expectation of 2.84 $\pm$ 0.15 (5.3\%). The final uncertainty is balanced between systematic and statistical components. The transverse momentum differential distribution of the Higgs is presented in Fig.~\ref{FidXS} (left). All results maintain model independence by floating the Higgs boson's branching ratio in four leptons, enhancing sensitivity to potential BSM effects in the decay.

The $\textnormal{H}\rightarrow\gamma\gamma$ fiducial analysis~\cite{FidXSHgaga} provides a comprehensive range of both inclusive and differential measurements to fully characterize the decay channel. The inclusive cross section is determined within dedicated phase-space regions designed to loosely target specific production modes. Additionally, the inclusive measurement is provided in VBF-enriched phase space, with several jet-related differential observables measured in this specific region. Further exploration of the phase space involves the measurement of some double differential observables, as illustrated in Fig.~\ref{FidXS} (right). The measured inclusive cross section:

\begin{center} 
$\sigma_{\textnormal{fid}}^{\textnormal{H}\gamma\gamma} = 73.4_{-5.3 (7.2\%)}^{+6.1(8.3\%)}$ fb = $73.4_{-5.3 (7.2\%)}^{+5.4(7.4\%)}$ (stat.) $_{-2.2 (2.3\%)}^{+2.4(3.3\%)}$ (sys.) fb
\end{center} 
closely aligns with the SM expectation of 75.4 $\pm$ 4.1 (5.4\%). The measured value's precision is nearly as expected, and the systematic component is well below the theoretical precision.

\begin{figure}[ht]
\centering
\includegraphics[width=11cm]{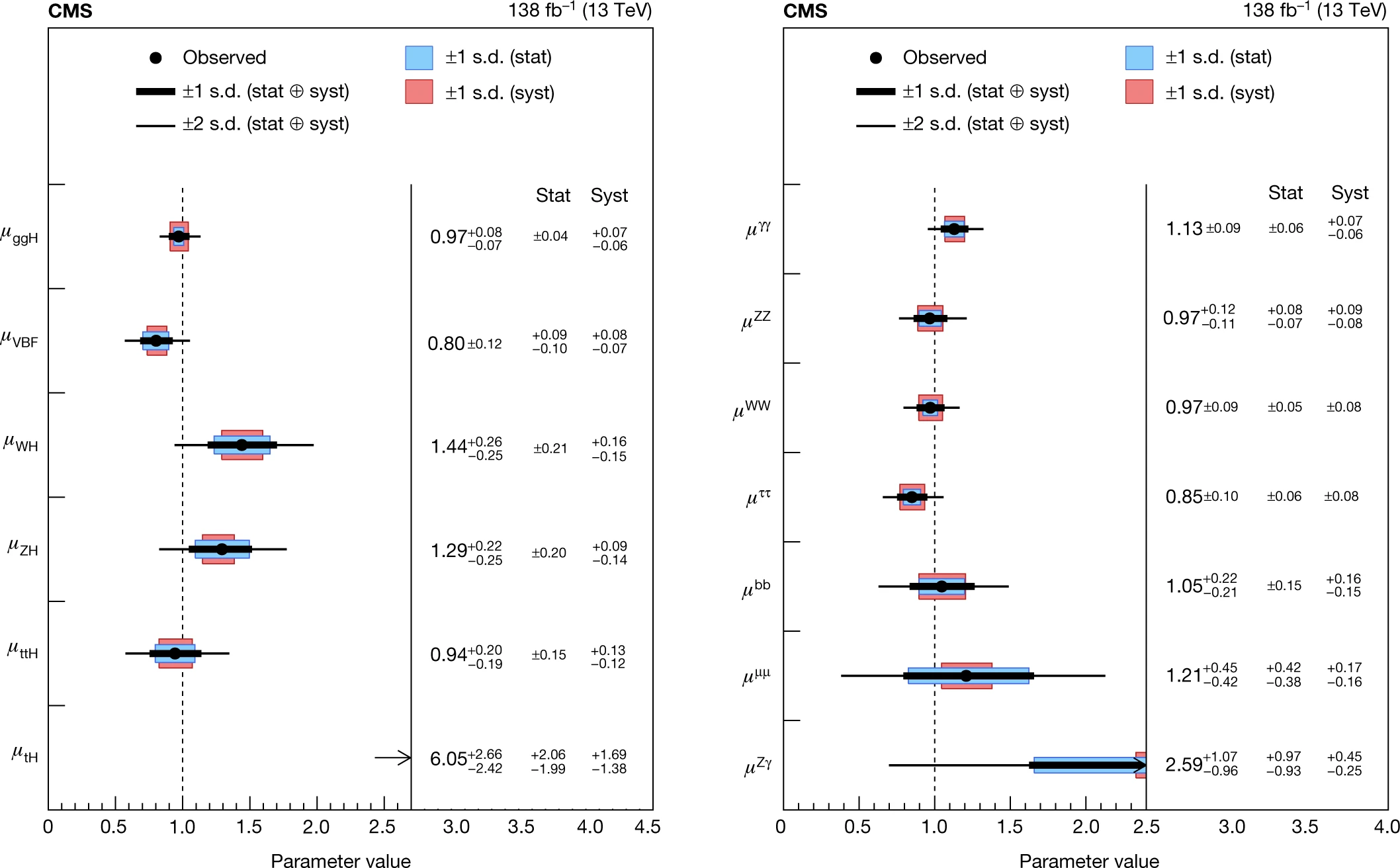}
\caption{Signal strength modifiers for production $\mu_{i}$ (left) and decay, $\mu_{f}$ (right) of the Higgs boson. The vertical dashed line at 1 represents the SM expectation.~\cite{portrait}.}
\label{decay_Prod}
\end{figure}

\section{Production and decay rate measurements}

From the time of the Higgs boson discovery, one of the most direct methods to assess the agreement between data and SM expectations involves the measurement of signal strength modifiers. These modifiers represent the scalings of the production and/or decay rates relative to the SM predictions. The most general compatibility test with the SM can be performed through the introduction of an inclusive signal strength modifier $\mu = (\sigma\mathcal{B})_{\textnormal{obs}}/(\sigma\mathcal{B})_{\textnormal{exp}}$, scaling both production and decay rates simultaneously~\cite{portrait}. The inclusive signal strength modifier measured for the Higgs boson to be 
\begin{center} 
$\mu = 1.002 \pm 0.057 = 1.002 \pm 0.036$ (theory) $\pm$ 0.033 (exp.) $\pm$ 0.029 (stat.),
\end{center}
showcasing a fourfold improvement in precision compared to the measurement done at the time of discovery ($\mu = 0.87 \pm 0.23$). Notably, the statistical uncertainty is now at a comparable level with theoretical and systematic uncertainties.

A more nuanced characterization of the Higgs boson profile emerges by introducing separate signal strength modifiers for production mechanisms ($\mu_{i}$) and decay channels ($\mu^{f}$). Results illustrated in Fig.~\ref{decay_Prod} reveal that all production modes, excluding tH, are observed with a significance of 5 standard deviations or higher. Overall, all the results show a good agreement with SM predictions, yielding p-values of 3.1\% and 30.1\% for $\mu_{i}$ and $\mu^{f}$, respectively.

\begin{figure}[ht]
\centering
\includegraphics[width=10cm]{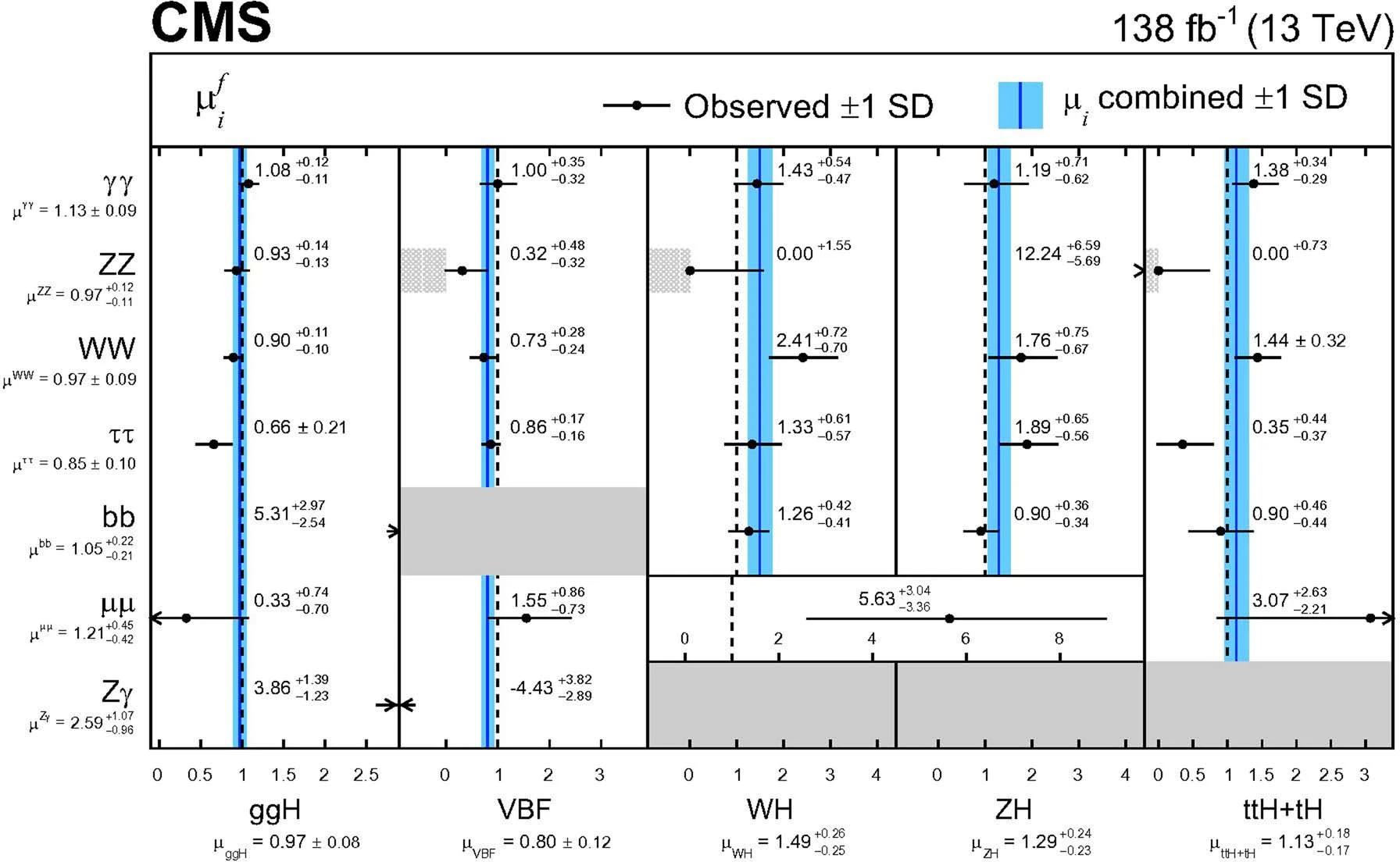}
\caption{Combined signal strenght modifier $\mu_{i}^{f}$ for all production and decay channels. The hatched gray bands indicate that the signal strength is forced to be positive and the dark grey shading indicates the absence of measurement. The vertical dashed line at 1 represents the SM expectation~\cite{portrait}.}
\label{sigStrength}
\end{figure}

For a more granular comparison of Higgs boson production and decay rates with the SM, assumptions on signal strength modifiers are further relaxed. Independent parameters (($\mu_{i}^{f}$) are introduced for each production channel and decay mode, as depicted in Fig.~\ref{sigStrength}. Once again, the results showcase an overall good agreement with SM predictions, corresponding to a p-value of 5.8\%.

\section{Coupling strength measurements}
\label{Sec:Coupling}
\begin{figure}[ht]
\centering
\includegraphics[width=8.1cm]{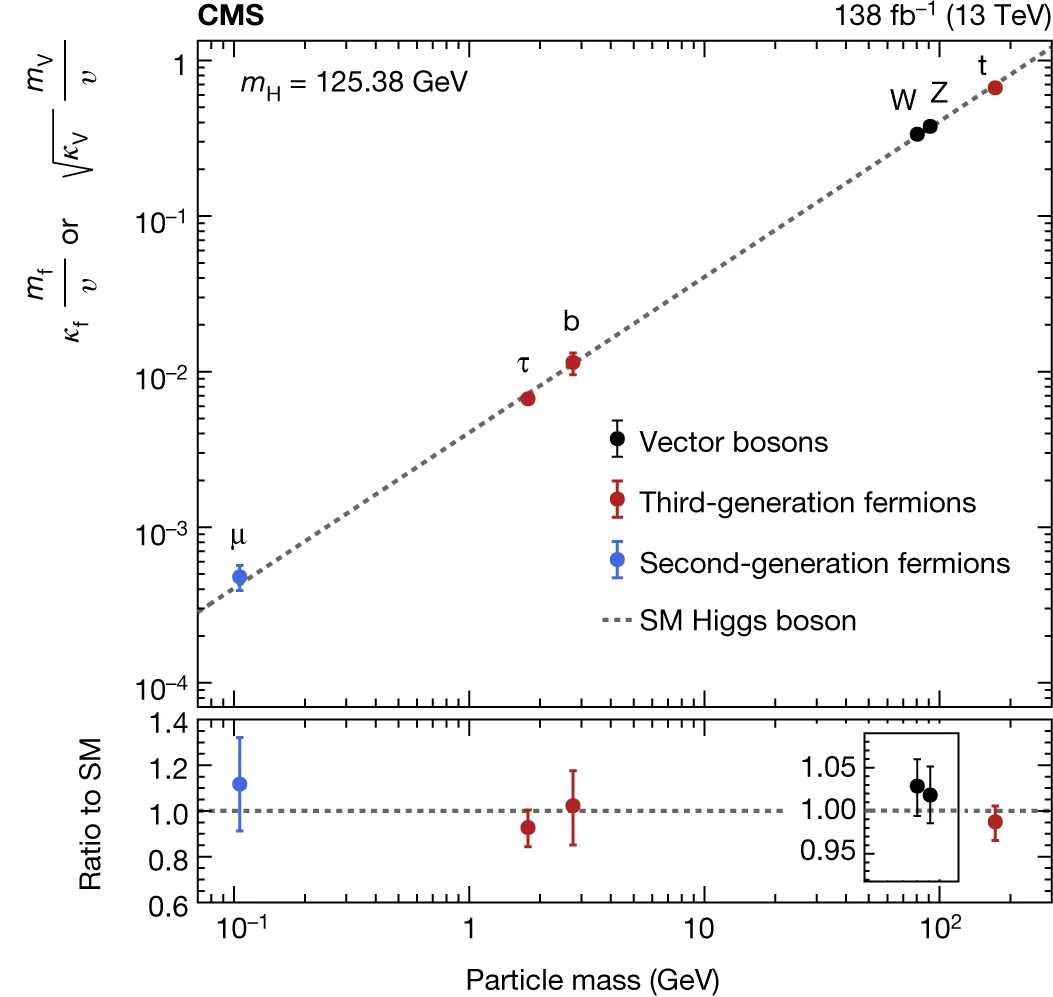}
\qquad
\qquad
\includegraphics[width=6.6cm]{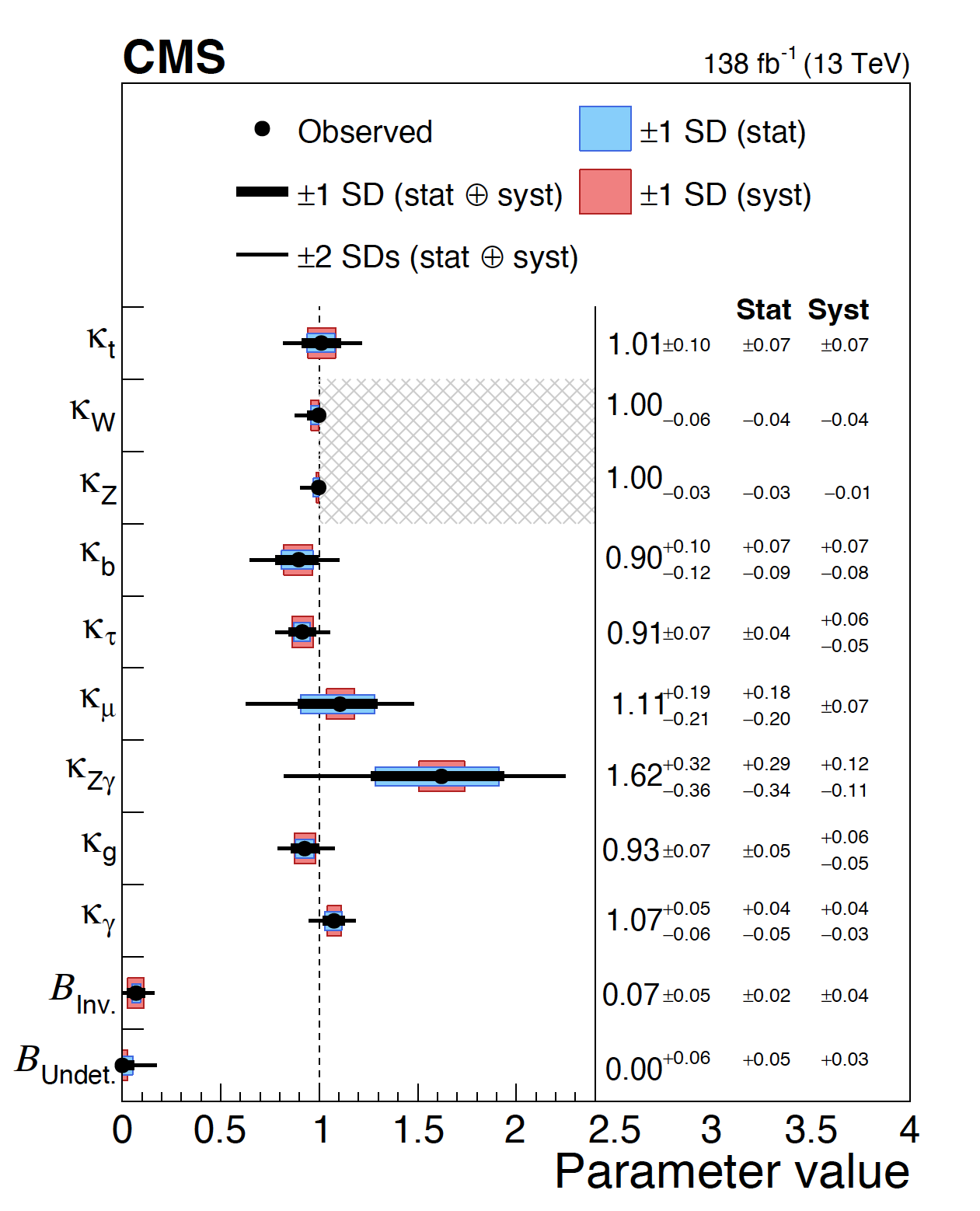}
\caption{Left: Measured modifiers of the Higgs boson couplings to vector bosons and fermions
as a function of the mass of the particles. Right: Coupling modifiers measured including loop-induced Higgs boson interactions. Invisible and undetected decay modes are considered in the fit.~\cite{portrait}.}
\label{Coupling}
\end{figure}

The characterization of the Higgs boson discussed in the previous section can be complemented by assessing its coupling to vector bosons and fermions in the so-called $k$-framework, employing coupling modifiers ($k_{i}$) as scaling factors for Higgs boson interactions. A global analysis of Higgs boson interactions introduces two parameters, scaling the inclusive couplings to vector bosons ($k_{V}$) and fermions ($k_{f}$). Stress testing the SM involves measuring the couplings of the Higgs boson with individual fermions and vector bosons, aligning well with SM predictions (dashed line in the left panel of Fig.~\ref{Coupling}). With the precision of these measurements, achieved with the substantial Run 2 dataset, couplings to second generation fermions started to be probed in more detail, while all other couplings are measured with remarkable precision, maintaining statistical and systematic uncertainties at similar levels.

To assess potential BSM effects, modifiers are introduced for effective couplings of the Higgs boson to gluons ($k_{g}$), photons ($k_{\gamma}$), and Z$\gamma$ ($k_{Z\gamma}$). The p-value for compatibility with the SM is 28\%, with couplings probed at the level of 10\%, except for $k_{\mu}$ and $k_{b}$ (20\%), and $k_{Z\gamma}$ (40\%). An additional degree of freedom is incorporated into the fit to account for invisible and undetected decay modes, restricting $k_{Z}$ and $k_{W}$ to be 1.0 or smaller. The results, shown in the right plot of Fig.~\ref{Coupling}, are compatible with the SM with a p-value of 33\%. Branching fractions to invisible and undetected Higgs boson decays are found to be compatible with zero, with 95\% CL upper limits of $\mathcal{B}_{\textnormal{Undet.}} <$ 0.16 and $\mathcal{B}_{\textnormal{Inv.}} <$ 0.17.

\subsection{The Higgs boson self-coupling}
\begin{figure}[ht]
\centering
\includegraphics[width=11cm]{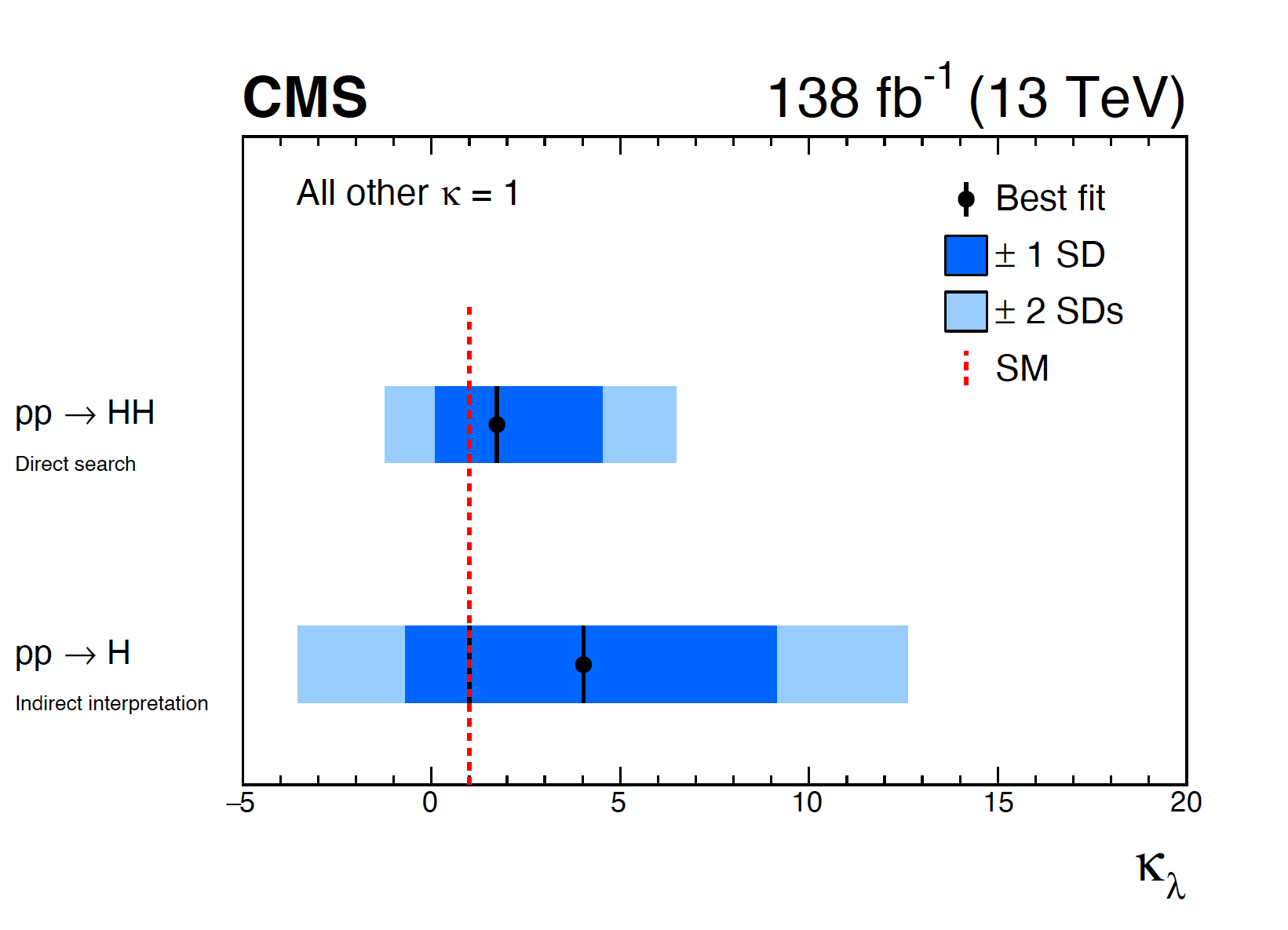}
\caption{Constraints on the Higgs boson self-coupling obtained directly from Higgs boson pair production and indirect interpretation of single H measurements.~\cite{portrait}}
\label{Self_Coupling}
\end{figure}

The measurement of the Higgs boson self-coupling is pivotal for understanding the properties of the Electroweak Symmetry Breaking (EWSB) mechanism in the SM. The strength of the Higgs boson self-interaction is expressed as $\lambda = m_{H}^{2}/(2\nu^{2})$, where $\nu$ is the vacuum expectation value of the Higgs field. Similar to the $k$-framework, this analysis establishes limits on the coupling modifier for the Higgs boson self-interaction ($k_{\lambda}$). Typically, constraints on $k_{\lambda}$ are derived from measurements of Higgs boson pair production, but this analysis extends the approach to single Higgs boson production and decay, which is sensitive to next-to-leading-order quantum corrections due to $k_{\lambda}$ modifications and can be used to set constraints complementary to those from searches for Higgs boson pair production. Both cases show the consistency of $k_{\lambda}$ with 1, as illustrated in Fig.~\ref{Self_Coupling}.

\section{Exotic decays of the Higgs boson}
The discovered Higgs boson demonstrates good compatibility with the Standard Model (SM), but ongoing exploration is underway for exotic decays as suggested by various Beyond Standard Model (BSM) theories. Notably, the ATLAS and CMS experiments have set upper bounds of 12\% and 16\%, respectively, on the branching fraction of the Higgs boson to new particles at a 95\% CL using Run 2 LHC data~\cite{portrait, exotic_Atlas}. Given these limitations, it is essential to persist in data analysis to search for direct evidence of new particles interacting with the Higgs boson and to scrutinize potential extensions of the SM of particle physics. In this section, various searches for the exotic decays of the Higgs boson are presented.

\subsection{Search for Higgs boson decays to a Z boson and a light pseudoscalar}
\begin{figure}[ht]
\centering
\includegraphics[width=10cm]{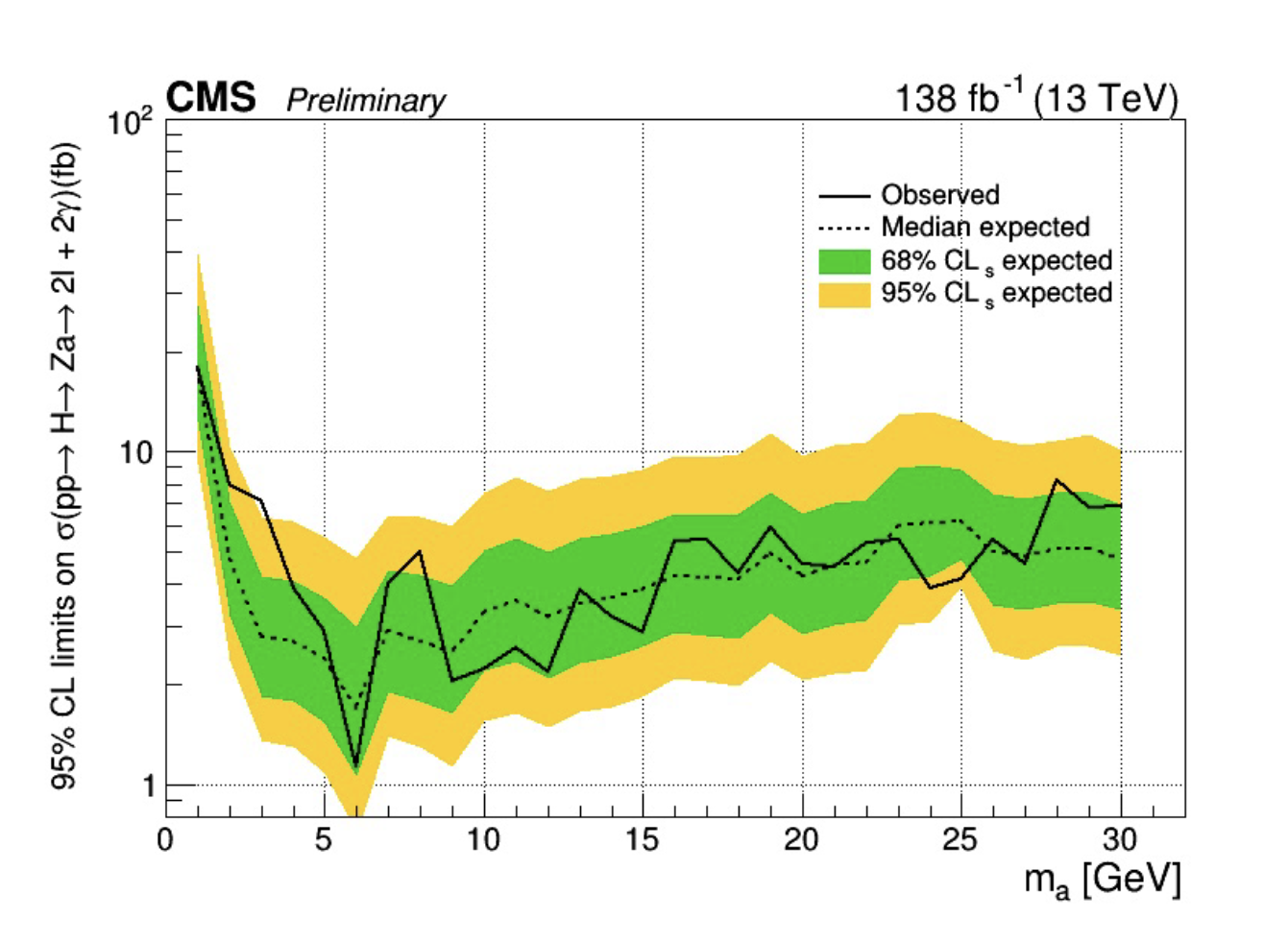}
\caption{Expected and observed 95\% CL limits on $\sigma(pp\rightarrow\textnormal{H})$ $\times \mathcal{B}(\textnormal{H}\rightarrow\textnormal{Z}a\rightarrow ll\gamma\gamma$) for $m_{a}$ from 1 to 30 GeV~\cite{exotic_Za}}
\label{exotic_Za}
\end{figure}

Motivated by theoretical models involving axion-like particles (ALPs) formulated within an effective field theory framework coupling ALPs to SM particles, a search for the Higgs boson decay process $\textnormal{H}\rightarrow\textnormal{Z}a\rightarrow ll\gamma\gamma$ (with $l = e, \mu$) is undertaken for the pseudoscalar, $a$, having masses ranging from 1 to 30 GeV~\cite{exotic_Za}. This study assumes the on-shell condition for the Z boson and excludes scenarios where the pseudoscalar has a mass below 1 GeV due to the potential instrumental merging of the two photons from the pseudoscalar decay. This specific situation requires a dedicated reconstruction technique and is therefore not addressed in the current analysis. The chosen final state provides a distinctive and experimentally clean signature, characterized by a low cross section from SM processes, marking the first such search at the LHC. Outcomes are derived through simultaneous unbinned maximum likelihood fits of the invariant mass of the four objects ($m_{ll\gamma\gamma}$), categorized based on the output of a BDT discriminator. As depicted in Fig.~\ref{exotic_Za}, no substantial deviation from the background-only hypothesis is observed, and upper limits at a 95\% CL are established for the production cross section times branching fraction.

\subsection{Searches for Higgs boson decays to a pair of light pseudoscalars}
The Two-Higgs-Doublet Models extended by an additional scalar singlet (2HDM+S) postulate the existence of seven physical scalar and pseudoscalar particles. Among these, one scalar aligns with the discovered Higgs boson with a mass of 125 GeV. In 2HDM+S, four coupling scenarios can avoid flavor-changing neutral currents at the tree level. This includes the next-to-minimal supersymmetric extension of the SM as a specific case. These models predict the decay of the Higgs boson into a pair of light pseudoscalar bosons whenever kinematically feasible, a phenomenon that remains unexcluded to date.
\subsubsection{Final state with four boosted photons}

\begin{figure}[ht]
\centering
\includegraphics[width=16cm]{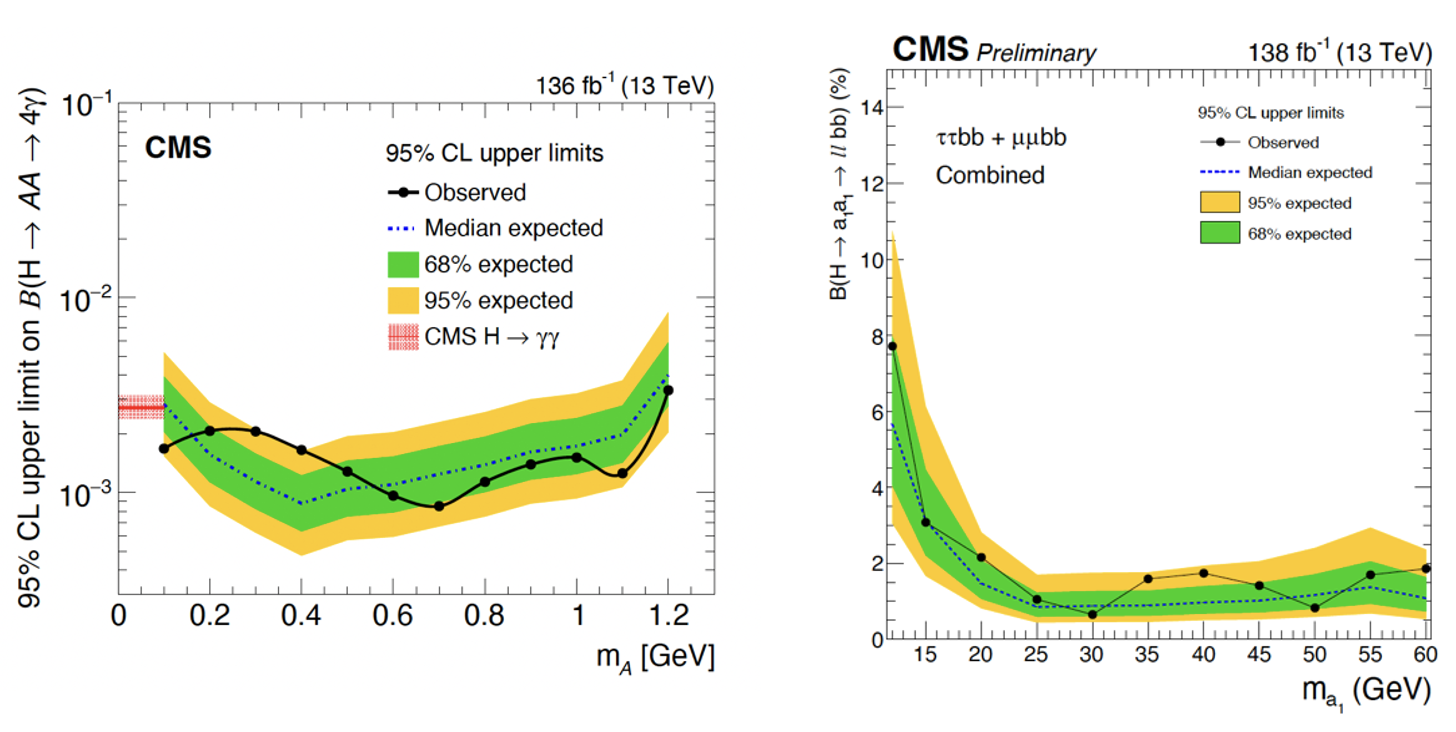}
\caption{Expected and observed 95\% CL limits on $\mathcal{B}(\textnormal{H}\rightarrow aa\rightarrow4\gamma$) from the merged diphoton analyses~\cite{exotic_aa1} (left), and on $\mathcal{B}(\textnormal{H}\rightarrow aa\rightarrow llbb$) under the 2HDM+S scenarios~\cite{exotic_aa2} (right)}
\label{exotic_aa}
\end{figure}

In scenarios of fermiophobic $a$ decays, where the branching fraction $\mathcal{B}(a\rightarrow\gamma\gamma)$ can approach unity, a search for $\textnormal{H}\rightarrow aa\rightarrow 4\gamma$ is conducted for a lower pseudoscalar mass range of 0.1 to 1.2 GeV~\cite{exotic_aa1}. The primary challenge in this search arises from the boosted pseudoscalars, leading to the reconstruction difficulty of the diphoton mass. Due to the boosted nature, the two photons are predominantly reconstructed as a single photon-like object ($\Gamma$), resulting in a degenerate measured invariant mass $m_{\Gamma\Gamma}$ peak with the SM $\textnormal{H}\rightarrow \gamma\gamma$ background. To address this challenge, a novel machine learning reconstruction algorithm is employed. This algorithm is trained on the electromagnetic calorimeter (ECAL) energy deposit patterns to measure the invariant mass $m_{\Gamma}$ of the merged diphoton candidates, a task deemed impossible with the standard CMS reconstruction algorithm. Discrimination between signal and background, comprised of the SM $\textnormal{H}\rightarrow \gamma\gamma$ process and other nonresonant processes, relies on the invariant masses of the two merged diphoton candidates. No observable excess of events above the estimated background is identified, and upper limits at a 95\% CL on $\mathcal{B}(\textnormal{H}\rightarrow aa\rightarrow4\gamma$) are established, as illustrated on the left side of Fig.~\ref{exotic_aa}.

\subsubsection{Final states with two b jets and two muons or two tau leptons}
A search for $\textnormal{H}\rightarrow aa$ is performed in the $\tau\tau bb$ or $\mu\mu bb$ final states, targeting pseudoscalar masses from 15 to 60 GeV~\cite{exotic_aa2}. The $\tau\tau bb$ channel benefits from the large branching fractions of pseudoscalars decaying into heavy quarks and leptons. Three $\tau\tau$ final states are considered ($e\mu$, $e\tau_{h}$, $\mu\tau_{h}$), with the $\tau_{h}\tau_{h}$ channel excluded due to high trigger thresholds. Event categorization involves b jet multiplicity, and deep neural networks aid signal-background discrimination. 

In the $\mu\mu bb$ final state, it benefits from the excellent $\mu\mu$ mass resolution and large pseudoscalar branching to heavy quarks. Employing event categorization based on jet transverse momentum, b jet tagging score, and invariant masses, discriminates signal from background. An unbinned maximum likelihood fit is performed on $\mu\mu$ invariant mass distribution to extract the results. No excess of events above the background-only expectation is observed in both final states. The combined analysis enhances sensitivity to potential exotic decays of the Higgs boson. Upper limits are set at 95\% CL on branching fractions in the 2HDM+S interpretation, as illustrated in Fig.~\ref{exotic_aa} (right).

\subsection{Invisible decays of the Higgs boson}
\begin{figure}[ht]
\centering
\includegraphics[width=8.5cm]{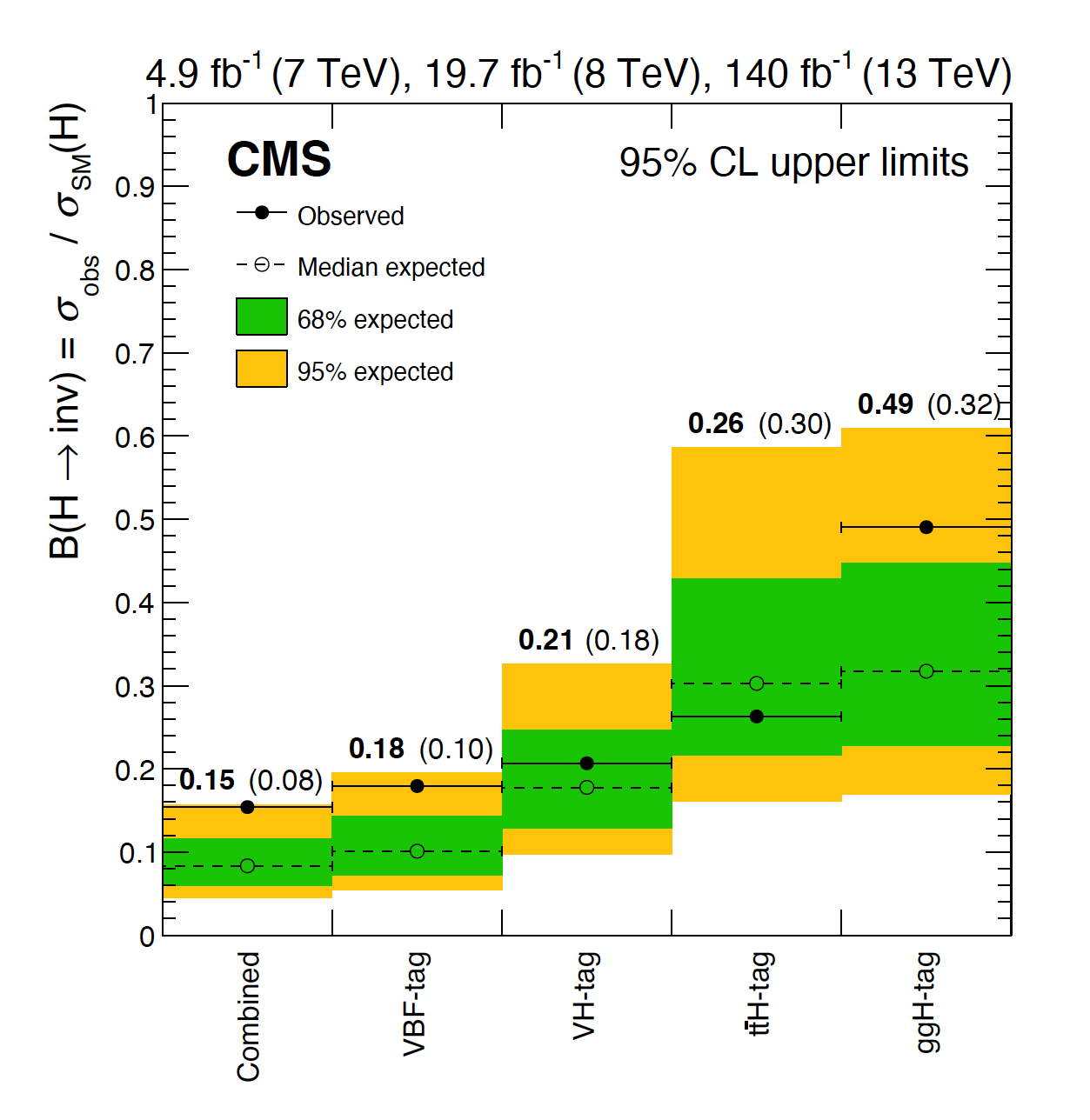}
\caption{Expected and observed 95\% CL limits on $\mathcal{B}(\textnormal{H}\rightarrow inv$ with sensitivity breakdown in production modes, combining data from Run 1 and Run 2~\cite{INV}.}
\label{HTinv}
\end{figure}

In SM, the predicted branching fraction $\mathcal{B}(\textnormal{H}\rightarrow inv$) is approximately 0.1\% through the $\textnormal{H}\rightarrow ZZ^{*}\rightarrow 4\nu$ channel. However, in BSM scenarios, particularly those involving a scalar sector connecting the SM and dark sectors, the Higgs boson may exhibit an increased rate by decaying into a pair of dark matter particles when kinematically feasible. Consequently, the pursuit of direct searches for $\textnormal{H}\rightarrow inv$ is crucial, and an observation of a higher-than-expected rate would serve as evidence for BSM physics. A recent search focuses on the invisible decays of the Higgs boson, produced in association with a pair of top quarks or a W/Z boson, leading to a fully hadronic final state is presented here~\cite{INV}. The signal is characterized by a large missing transverse momentum from the invisibly decaying Higgs boson, recoiling from its associated hadronically decaying particles. The analysis reveals no significant excess of events beyond the SM prediction. At a 95\% CL, an upper limit of 0.47 (0.40 expected) is imposed on the branching fraction $\mathcal{B}(\textnormal{H}\rightarrow inv$), assuming the SM production cross section of the Higgs boson. Combining this analysis with previous searches in complementary production modes at $\sqrt{s}$= 7, 8, and 13 TeV in Run 1 and Run 2 data of the LHC establishes a 95\% CL upper limit on $\mathcal{B}(\textnormal{H}\rightarrow inv$) of 0.15 (0.08 expected), as depicted in Fig.~\ref{HTinv}.

\section{Searches for double Higgs boson production}

\begin{figure}[ht]
\centering
\includegraphics[width=16cm]{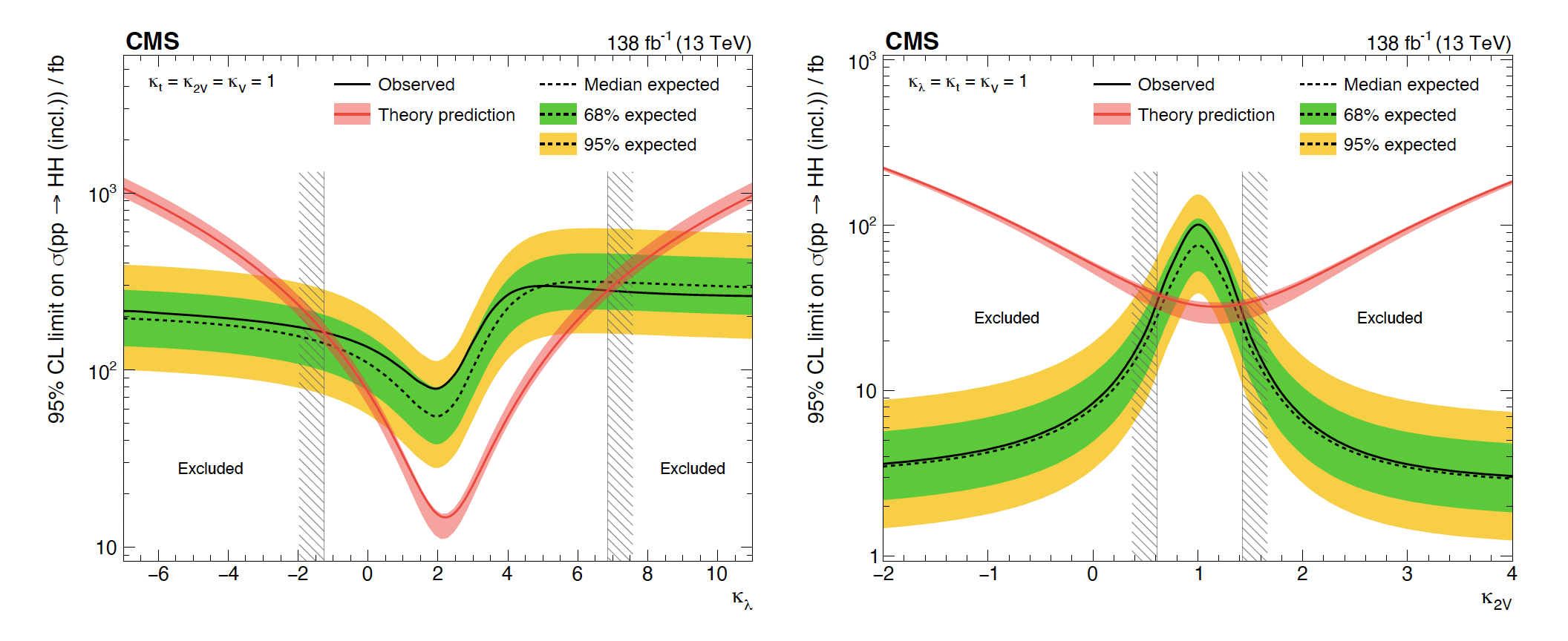}
\caption{Combined expected and observed 95\% CL upper limits on the HH production cross section for different values of $k_{\lambda}$ (left) and $k_{2V}$ (right)~\cite{portrait}.}
\label{HH}
\end{figure}

The study of double Higgs boson production provides an avenue to measure the Higgs boson self-coupling (see section~\ref{Sec:Coupling}) and the coupling between two Higgs bosons and two vector bosons. To ease the interpretation of the results, the coupling values are usually presented in terms of their ratios with respect to the standard model expectations, denoted as $k_{\lambda}$ and $k_{2V}$. The main challenge lies in the significantly lower production cross section, three orders of magnitude less than single Higgs boson production, leading to results constrained by the statistical power of the dataset. The analysis focuses on final states where at least one Higgs boson decays into a pair of b quarks, profiting from the large $\textnormal{H}\rightarrow bb$ branching fraction (58\%). The combined analysis of various final states (bbZZ, multilepton, bb$\gamma\gamma$, bb$\tau\tau$, bbbb), illustrated in Fig.~\ref{HH}, constrains the parameter $k_{\lambda}$ within the interval -1.24 $< k_{\lambda} <$ 6.49 and the parameter $k_{2V}$ within the interval 0.67 $< k_{2V} <$ 1.38, with a noteworthy exclusion of $k_{2V}$ = 0 at a significance of 6.6 standard deviations~\cite{portrait}.

\section{Conclusions}
The journey to comprehend the Higgs boson's properties, initiated with its discovery in July 2012, has reached a significant milestone with the comprehensive analysis of the full Run 2 dataset. This extensive dataset enables the exploration and combination of a large number of production modes and decay channels, facilitating an intricate characterization of the Higgs boson's profile. Precise measurements of its production cross sections, production and decay rates and couplings to other particles underscore the remarkable agreement with Standard Model expectations. Notably, no indications of new physics beyond the Standard Model are observed, as evident from the consistent and negligible branching fraction for Higgs to invisible decays. Several results in this analysis showcase a substantial enhancement in precision compared to the corresponding results at the time of the Higgs boson discovery. The full Run 2 dataset marks an entrance into the precision physics realm, where many experimental results are starting to have comparable statistical and systematic uncertainties, soon becoming predominantly limited by the latter.

\end{document}